\documentclass[12pt]{article}

\newcommand{\eq}{\begin{equation}}
\newcommand{\eqx}{\end{equation}}
\newcommand{\eqn}{\begin{eqnarray}}
\newcommand{\eqnx}{\end{eqnarray}}
\newcommand{\f}[2]{\frac{#1}{#2}}

\newcommand{\tr}{\mbox{\rm tr}\,}

\newcommand{\Lm}{\Lambda}
\newcommand{\dl}{\delta}
\newcommand{\ff}{{\cal F}}
\newcommand{\nn}{{\cal N}}
\newcommand{\der}[2]{\f{\partial{#1}}{\partial{#2}}}
\newcommand{\Qt}{\tilde{Q}}

\title{Effective matter superpotentials from Wishart random matrices.\footnote{Work partially
supported by the European Commission RTN programme HPRN-CT-2000-00131.}\\}

\author{Yves Demasure$^{a,b}$\footnote{e-mail: {\tt demasure@nordita.dk}}\
\ and Romuald A. Janik$^{c,d}$\footnote{
e-mail: {\tt janik@nbi.dk}}\\ \\
$^a$ Instituut voor Theoretische Fysica, Katholieke Universiteit Leuven,\\
Celestijnenlaan 200D, B-3001 Leuven, Belgium\\
$^b$ NORDITA,\\
Blegdamsvej 17, DK-2100 Copenhagen,\\
Denmark\\
$^c$ The Niels Bohr Institute,\\
Blegdamsvej 17, DK-2100 Copenhagen,\\
Denmark\\
$^d$ Jagellonian University,\\
Reymonta 4, 30-059 Krakow,\\
Poland}

\begin{document}

\begin{titlepage}

\rightline{KUL-TF-2002-17}
\rightline{NORDITA-HE-2003-14}
\setcounter{page}{0}

\vskip 1.6cm

\begin{center}

{\LARGE Effective matter superpotentials from Wishart random matrices.}
\footnote{Work partially supported by the European Commission RTN 
programme HPRN-CT-2000-00131.}

\vskip 1.4cm

Yves Demasure$^{a,b}$\footnote{e-mail: {\tt demasure@nordita.dk}}\
\ and Romuald A. Janik$^{c,d}$\footnote{
e-mail: {\tt janik@nbi.dk}}\\ 
\vskip 0.3cm
 
{\small
$^a$ Instituut voor Theoretische Fysica, Katholieke Universiteit Leuven,}\\
{\small Celestijnenlaan 200D, B-3001 Leuven, Belgium}\\
{\small $^b$ NORDITA,}\\
{\small Blegdamsvej 17, DK-2100 Copenhagen, Denmark}\\
{\small $^c$ The Niels Bohr Institute,}\\
{\small Blegdamsvej 17, DK-2100 Copenhagen, Denmark}\\
{\small $^d$ Jagellonian University,}\\
{\small Reymonta 4, 30-059 Krakow, Poland}

\end{center}

\vskip 1.4cm

\begin{abstract}
We show how within the Dijkgraaf-Vafa prescription one can derive
superpotentials for matter fields. The ingredients forming the
non-perturbative Affleck-Dine-Seiberg superpotentials arise from
constrained matrix integrals, which are equivalent to classical
complex Wishart random matrices. The mechanism is similar to the way
the Veneziano-Yankielowicz superpotential arises from the matrix model
measure.
\end{abstract}
\end{titlepage}

\section{Introduction}

An interesting problem of the nonperturbative dynamics of
$\nn=1$ supersymmetric gauge theories is the study of gaugino
condensation. This is described by an effective superpotential
$W_{eff}(S)$ where $S$ is the glueball superfield $S=-(1/32\pi^2) \tr
W_\alpha W^\alpha$ whose lowest component is the gaugino bilinear.

Recently a completely novel approach was developed for calculating
these superpotentials for theories with matter fields $\Phi$ in the adjoint
representation and an arbitrary tree-level superpotential $W_{tree}(\Phi)$.
Initially formulated in the framework
of geometric transitions \cite{V1,V2} it was subsequently translated
into a problem in random matrix theory \cite{DV1,DVP,DVP2}.
This generated a lot of work
\cite{DOREY,FERRARI,FUJI,BERENSTEIN,GORSKY,FERRETI,MCGREEVY}
culminating in a field 
theoretical proof of the perturbative part of the conjecture \cite{PROOF}.
Subsequently these ideas were extended to matter fields in the fundamental
representation \cite{FERRETI,MCGREEVY} where $W_{eff}(S)$ was
calculated for $N_f$ flavors interacting with the adjoint field.
In the above cases the effective superpotential is expressed purely as a
function of $S$.

The object of this note is to show that the effective superpotentials for
matter fields can be calculated directly within the same framework in a natural
manner. This is interesting as the $N_f$-$N_c$ phase diagram of $\nn=1$
supersymmetric $U(N_c)$ gauge theories with fundamental matter has a very rich
structure.  
Therefore it would be nice to be able to address
these issues also within the Dijkgraaf-Vafa framework.
The appropriate constrained random matrices relevant in this
setting are the so-called (complex) Wishart random matrices
(\cite{WISHART}, for some recent developments see \cite{US}).

The plan of this paper is as follows. In section 2 we will briefly
recall the Dijkgraaf-Vafa prescription for theories with matter fields
in both the fundamental and adjoint representation. In section 3 we
will show how to calculate the mesonic superpotential purely in the
random matrix framework. We find that a qualitative difference in the
random matrix setup appears for $N_c<N_f$.
We briefly discuss some
generic features of $N_f=N_c$, the full solution of which remains,
however, an open problem. We close the paper with a discussion.

\section{Dijkgraaf-Vafa proposal with matter fields}

The Dijkgraaf-Vafa proposal was originally formulated just for theories
with adjoint matter. It has subsequently been
realized\footnote{Although already in \cite{DVP} hints about the
treatment of fundamental matter appear.} \cite{FERRETI,MCGREEVY} that
a natural extension of the proposal to theories with matter fields in
the fundamental representation is
\eq
\label{e.prop}
W_{eff}(S)=-N_c S \log \f{S}{\Lm^3} -2\pi i \tau_0 S + N_c
\der{\ff_{\chi=2}}{S}+\ff_{\chi=1}(S)
\eqx
where the $\ff_i$'s are defined through a matrix integral
\eq
\label{e.presc}
e^{-\sum_{\chi} \f{1}{g_s^\chi} \ff_\chi (S)} =
\int D\Phi DQ_i D\Qt_i \exp\left\{ -\f{1}{g_s}
W_{tree}(\Phi,Q_i,\Qt_i) \right\}
\eqx
Here $\Phi$ is an $N\times N$ matrix, while the $Q_i$'s are $N$
component vectors. In this expression the limit $N\to \infty$, $g_s
\to 0$ and $g_s N=S=const$ is understood\footnote{In this paper we
consider only the case of no gauge symmetry breaking and thus only a
single $S$ superfield.}.
In (\ref{e.prop}) the the first term, being the
Veneziano-Yankielowicz effective potential \cite{VENYANK}, can equivalently
be obtained from the prescription (\ref{e.presc}), through the inclusion
of the correct volume factor in the measure for the random matrix model 
\cite{DVP}:
\eq
\frac{1}{Vol( U (N) )} \int D\Phi
\eqx
The volume factor is  proportional to $\prod_{k=1}^{N-1}  k!$.
When $N\to \infty$ this behaves like $\exp(\f{1}{2} N^2 \log N)$.
This factor will then contribute to $\ff_2(S)$ through the relation
\eq
-\f{N^2}{S^2} \ff_2(S) =  \f{1}{2} N^2 \log N +\ldots
\eqx
So $\ff_2 \sim -\f{1}{2} S^2 \log S$ where we transformed the $N$ into
$S$ through $S= g_s N$. The remaining factor of $g_s$ is removed by
the logarithmic part of the random matrix free energy.  
After differentiation, $N_c\partial\ff_2/\partial S$ gives the
$-N_c S\log S$ Veneziano-Yankielowicz term.

>From prescription (\ref{e.prop}), one obtains the effective superpotential for
the glueball superfield $S$ only. In the case of fundamental matter
fields there are a number of other possible gauge invariant and holomorphic
polynomials in the
superfields like the mesonic fields $X_{ij}=Q^\dagger_i \Qt_j$.
We would like to show how to naturally incorporate these degrees of freedom
in the DV framework and how to obtain superpotentials for these
fields. As a check of 
the method we will recover the standard Affleck-Dine-Seiberg
superpotential \cite{ADS} for $N_f<N_c$
\eq
(N_c-N_f) \left( \f{\Lm^{3N_c-N_f}}{\det X} \right)^{\f{1}{N_c-N_f}}
\eqx
We would like to stay purely within the random matrix framework
and thus we refrain from performing `integrating in' and other field
theoretical procedures.

\section{Superpotential for the mesonic superfields}

First let us note that it is natural to expect that the proposal
(\ref{e.prop}) should also hold for theories with only fundamental
matter fields. This can be justified by the argument that one can always
integrate out the adjoint field.

Now we want to express the effective superpotential in terms of
mesonic superfields $X_{ij}=Q^\dagger_i \Qt_j$. The prescription that we want
to advocate is to perform only a partial integration over the $Q's$ in
(\ref{e.presc}) and impose the constraint $X_{ij}=Q^\dagger_i \Qt_j$
directly in the matrix integral i.e.
\eq
e^{-\sum_\chi \left(\f{N}{S}\right)^{\chi} \ff_\chi (S,X)} =
\int  DQ_i D\Qt_i \dl(X_{ij}- Q_i^\dagger \Qt_j) \exp\left\{ -\f{N}{S}
W_{tree}(Q_i,\Qt_i) \right\}
\eqx
where we made the substitution $g_s=S/N$. For $N_f<N_c$,
$W_{tree}(Q_i,\Qt_i)$ is an arbitrary holomorphic polynomial 
in $X$: $V_{tree}(X)$.  Then the effective
superpotential is obtained from
\eq
W_{eff}(S,X)=-N_c S \log \f{S}{\Lm^3} -2\pi i \tau_0 S + N_c
\der{\ff_{\chi=2}(S,X)}{S}+\ff_{\chi=1}(S,X)
\eqx
In the above formula the Veneziano-Yankielowicz term is essentially
put in by hand as required by gauge theory dynamics. A complete
understanding of its precise origin, in particular in the above
situation without the adjoint field, remains one of the few missing
pieces of the DV proposal.
The tree level superpotential $V_{tree}(X)$ contributes directly to
$\ff_{\chi=1}(S,X)$. An additional contribution to $\ff_{\chi=1}(S,X)$
will come from the constrained integral over $N_f$ vectors of length $N$
\eq
\label{e.meas}
\Lm^{-2N_f N}\int  DQ_i D\Qt_i \dl(X_{ij}- Q^\dagger_i \Qt_j)
\eqx
where the factor $\Lm^{-2N_f N}$ was included in order to keep the integration
measure dimensionless.
Up to an inessential term $\exp(-\tr X)$ this is just the probablity
distribution of (complex) Wishart random matrices \cite{WISHART}. In fact it
depends crucially on the relative magnitude of $N_f$ and $N$ ($\sim N_c$).
If $N_f \leq N$ the answer is given by\footnote{See e.g. \cite{US} for
a quick proof and eq (15) in \cite{FYODOROV} for the numerical
coefficient $A$.}
\eq
A \cdot \Lm^{-2N_f^2 } \cdot \left( (\det X)/\Lm^{2N_f} \right)^{N-N_f}
\eqx
where
\eq
A=\f{(2\pi)^{\f{N(N+1)}{2}}}{\prod_{j=N-N_f+1}^N (j-1)!} \sim
\f{1}{(N!)^{N_f}} \sim e^{-N_f N \log N}
\eqx
For $N_f>N$ the situation changes drastically and a number of
constraints appear between the elements of $X$. This case has been
solved in \cite{US}. It is reassuring that
a qualitative change of the behaviour of the relevant matrix model
occurs when $N_f>N_c$ similarly as for $\nn=1$ supersymmetric gauge
theories \cite{SEIBERG}.

Performing exactly the same reasoning as for the volume of the random
matrix measure one gets a contribution to the superpotential
$\ff_1=N_f S\log S$. Thus the Veneziano-Yankielowicz term gets
transformed into
\eq
(N_c-N_f)S(-\log \f{S}{\Lambda^3} +a)
\eqx
It is convenient to choose the scale $\Lambda$ such that $a=1$.

The determinant gives a contribution to the $\ff_1$ term as $-S\log
\det X/\Lm^{2N_f}$ so the full superpotential for both $S$ and $X$ is
\eq
\label{e.weffsx}
W_{eff}(S,X)=(N_c-N_f)S(-\log \f{S}{\Lambda^3} +1) -S\log\left(\f{\det
X}{\Lm^{2N_f}}\right)  +V_{tree}(X)
\eqx
Note that the Intriligator-Leigh-Seiberg linearity principle
\cite{ILS} appears naturally within the matrix model formalism, as
already pointed out in \cite{FERRARI}. In the above case this is clear
as the tree level potential $V_{tree}(X)$ does not affect
the jacobian from the measure (\ref{e.meas}).
Solving the $F$ flatness condition
$\partial W_{eff}(S)/\partial S=0$ for $S$ gives
\eq
\log \f{S}{\Lm^3}=-\f{1}{N_c-N_f} \log\left(\f{\det X}{\Lm^{2N_f}}\right)
\eqx
Plugging this back into (\ref{e.weffsx}) we obtain the
Affleck-Dine-Seiberg superpotential which was generated by the measure
(\ref{e.meas}) in addition to the unmodified tree level $V_{tree}(X)$
\eq
W_{eff}(X)=(N_c-N_f) \left( \f{\Lm^{b_0}}{\det X} \right)^{\f{1}{N_c-N_f}}
+V_{tree}(X)
\eqx
with $b_0=3N_c-N_f$ the coefficient of the 1-loop $\beta$ function.
We find it quite surprising that the rather complicated form of the
nonperturbative Affleck-Dine-Seiberg superpotential arose just from
the large $N$ asymptotics of the constrained matrix measure (\ref{e.meas}).

Let us briefly comment on the case $N_f=N_c$ with $V_{tree}(X)=0$. From
the general form of (\ref{e.weffsx}) we see that the logarithmic term
vanishes and $S$ appears only {\em linearly} as
\eq
W_{eff}=S\log \left(\mbox{\rm `something'}/\Lm^\# \right)
\eqx
thus it generates in a natural way a constraint surface satisfying the
equation `something'$=\Lm^\#$. On this surface the effective potential
vanishes. This is the correct qualitative behaviour for $N_f=N_c$
\cite{SEIBERG}. However in order to make the above analysis precise
one would have to incorporate properly into the matrix formalism the
new holomorphic gauge invariants which arise for $N_f \geq N_c$
i.e. the baryonic fields (that is why we put `something' in the above
formula). We leave this very much non-trivial case for further
investigation \cite{WIP}. 

\section{Discussion}

In this short note we have shown how one can recover effective
superpotentials for matter fields in the fundamental representation
directly from the DV matrix model framework. The matter fields are
represented by vectors (more generally rectangular matrices). When
passing to gauge invariant variables 
one imposes appropriate constraints on the integration over the matter
fields. The large $N_c$ asymptotics of the constrained matrix integral
reproduces the whole nonperturbative Affleck-Dine-Seiberg effective
superpotential. 
In the case of $N_f<N_c$ this
is just the classical complex Wishart random matrix integral.
It would be very interesting to understand this also from the field
theoretical perspective, especially as it seems to be similar to the
way that the Veneziano-Yankielowicz superpotential arises from the
measure for the (adjoint) matrix model.
The latter case is not encompassed by the recent field theoretical
proof \cite{PROOF} of the perturbative part of the DV proposal.

The prescription adopted in this paper is unambigous for $N_f<N_c$ and
can be easily generalized to other cases like bifundamental matter,
$SO(N_c)$ gauge groups etc.
When $N_f=N_c$ we see that a `quantum' moduli space seems to arise
naturally but in order to say anything definite one should still
incorporate baryonic fields into the above framework. This is
nontrivial since the asymptotics of the matrix integrals are extracted
at large $N_c$ and the definition of the baryonic field becomes
ambiguous. We leave this open problem for future study.

\bigskip

\noindent{\bf Acknowledgments} RJ was supported by the EU
network on ``Discrete Random Geometry'' and KBN grant~2P03B09622.
YD was supported by an EC Marie Curie Training site Fellowship at
Nordita, under contract number HPMT-CT-2000-00010.

\medskip

\noindent{\bf Note added} As this paper was being completed ref.
\cite{SUZUKI} appeared, where related results were obtained using a
different method --- an integrating-in procedure for the
Dijkgraaf-Vafa superpotential $W_{eff}(S)$.

\end{document}